\documentclass[aps]{revtex4}
\usepackage{epsfig}

\begin{document}
\title{{\itshape Ordering dynamics in the presence of multiple phases}}

\author{Alberto Petri$^{\rm a}$, Miguel Iba\~nez de
Berganza$^{\rm b}$ and Vittorio Loreto$^{\rm c}$
\\\vspace{6pt}
 $^{\rm a}${\em{Istituto dei Sistemi Complessi - CNR, via del
 Fosso del Cavaliere 100, 0033 Roma, Italy}};
 $^{\rm b}${\em{Instituto de F\'isica  Te\'orica  (CSIC),
 Universidad Aut\'onoma de Madrid, Madrid 28047, Spain}}; $^{\rm
c}${\em{Dipartimento di Fisica, Universit\`a La Sapienza, P.le A.
Moro 2, 00185 Roma, Italy}} } 

\begin{abstract}

The dynamics of the 2D Potts ferromagnet when quenched below the
transition temperature is investigated in the case of
discontinuous phase transition, which is interesting for
understanding the  non equilibrium dynamics of systems with many
competing equivalent low temperature phases, that appears to be
not much explored. After briefly reviewing some recent findings,
we focus on the numerical study of quenches just below the
transition temperature on square lattices.  We show that, up to a
certain time, metastable states can be observed for which energy
stays constant above the equilibrium energy and the
self-correlation function displays a fast decay.

\end{abstract}
\maketitle

\section{Introduction}

The way in which thermodynamical systems attain equilibrium when
driven through a phase transition has been subject of much work
\cite{bray94,hoenberg77,gunton83,binder87}. In fact, while
equilibrium requires generally an infinite time to be reached,
different dynamical regimes can be observed which present often
peculiar properties (different kinds of dynamic scaling, critical
slowing-down, spatial self-affinity, etc.) depending on the
thermal history and the system. Dynamics of different phase
transitions have been widely investigated, e.g from a disordered
to an ordered phase, or between two ordered phases. Among the
model systems, the Ising model displays  both instances: in zero
magnetic field it undergoes a continuous phase transition from the
high temperature paramagnetic phase to the low temperature
ferromagnetic phase; the transition is discontinuous when at
constant low temperature an external magnetic field is reversed.
Both problems have been widely investigated. In the first case,
one of the main results  is contained in the Allen-Cahn law for
the ordered phase domain growth at low temperature, which predicts
a power law decay of the  energy  $\simeq t^{-1/k}$, with $k=2,3$
\cite{bray94,hoenberg77}. For what concerns the second case, the
metastability associated with the  field-driven first order phase
transition has been studied with nucleation theory and its
generalizations \cite{gunton83,binder87}, which describe
effectively the nucleation time and other features of the
metastable states \cite{langer67,binder87,rikvold94,rikvold02}.

Not many results seem instead available for the ordering dynamics
of systems possessing many equivalent ordered phases
\cite{bray94,schmittmann85}. On one hand, the coarsening at low temperature
presents peculiar properties not well described by the Allen-Cahn
theory.  This is predicted on general bases by the Lifshitz's
criterium \cite{lifshitz62}, according to which a system
possessing many equivalent ordered phases may equilibrate towards
non-homogeneous phases when quenched below its critical point if
their number is larger than the real space dimensionality. Thus,
equilibrium phases at low temperature can be not uniquely defined
and systems may display complex behaviours. On the other hand, and
differently from what happens for the Ising case, there exists
little theoretical study of the metastability near the
temperature-driven transition of the  $2D$ $q$-Potts model (which
is first order for $q>4$).  A recent one  \cite{meunier00}
 predicts in particular that, for this model, metastability is a
finite size effect. The study is based on a droplet expansion and,
to our knowledge, has not yet been related with the underlying
dynamics \cite{binder81} (metastability in the Potts model has
been investigated with different aims in \cite{berg04}, and for
the field driven transition in \cite{rutkevich02}).

The aim of the present paper is to resume some recent findings on
the dynamics of the $d=2$ Potts model \cite{wu82} and to present
some new results concerning metastability in the undercooled
states just below the transition. Investigation of temperature
driven metastability in this model seems particularly important
for several reasons. The model presents interest in itself, being
an effective model for systems with many possible broken
symmetries, and mimics the dynamics of a large variety of real
systems, such as gas adsorption \cite{wu82}, froth \cite{grest88},
grain growth \cite{thomas06}, biological cell
\cite{grest89,graner92}. On the other hand, as mentioned before,
the existence, or not, of metastable states associated to the
temperature-driven first order transition of the Potts model is an
old problem, firstly proposed by Binder \cite{binder81}, and which
remains still open.

The next section quickly recall the Potts model and presents a
brief review of some recent finding on its dynamics on the square
lattice in different temperature regions below the transition
temperature. After introducing  some elementary notions about
metastable states in thermodynamical systems and a brief
discussion on some recent findings concerning the metastable
states of the Potts model, Sec. 3 points out the presence of
metastable states for the model on square lattice simulations
 in terms of energy and self-correlation
function. Section 4 presents a short summary and perspectives.

\section{Potts model dynamics below the transition temperature}

The Potts model is described by a Hamiltonian of the kind
\cite{wu82}:

\begin{equation}
\label{potts} {\cal H}\lbrace \eta\rbrace =  \sum_{\langle ij
\rangle} (1-\delta_{\eta_i,\eta_j}),
\end{equation}
where $\eta_i$ is the state of site $i$ that can assume one out of
$q$ different values, usually identified by an integer: $\eta_i
\in [1,q]$. When $q=2$ Eq. (\ref{potts}) reduces  to the Ising
Hamiltonian (within multiplicative and additive constants). Here
the case of nearest neighbour interaction on square lattice will
be considered. It presents a discontinuous equilibrium temperature
driven phase transition at $T_q$ for $q>4$ \cite{wu82}.

The non equilibrium dynamics of the Potts model presents several
peculiar features. For example the determination of the domain
growth exponents in the Allen-Cahn law at low temperature remained
problematic for a long time \cite{grest88,derrida96},  whereas it
was just an artifact of the residual energy
\cite{deoliveira01,ferrero07} (see below). From a general point of
view, a recent survey on the different dynamical regimes displayed
by the Potts model has been done by Ferrero and Cannas
\cite{ferrero07b}. These regimes are observed in finite-size
realizations of the system when they are cooled below the
transition temperature $T_q$ (via Monte Carlo dynamics) and 
depend on the final temperature $T$ that, according to the dynamical 
features, can be grouped in
roughly three different ranges \cite{ferrero07b}, delimited by the
temperatures $0 < T_g < T^* <T_c < T_q$.

\begin{itemize}

\item For $T^* \alt T <T_c$ relaxation is dominated by simple
coarsening, which follows the quench according the Allen-Cahn law
\cite{bray94} for the growth of the domain size, withe energy per
site decaying as : $e(t) \simeq t^{-1/2}$.

\item For $T_g \alt T \alt T^*$ the simple coarsening is
interrupted at long time scales and the system is trapped with
finite probability in some configurations of high symmetry with
characterisitic lengths that grow with the system size
\cite{ferrero07b}. These states can be identified with those
predicted by Lifshitz, from which the system escapes by activated
dynamics.

\item For $T \lesssim T_g$ the system gets stuck in some kind of
frozen or glassy states with well defined lifetime, which diverges
with $T \rightarrow 0$ as $e^{c/T}$ \cite{ferrero07b}. An
interesting aspect \cite{ferrero07,ibanez07,deoliveira01b} is that
during relaxation the system obeys a generalized Allen-Cahn law
wihich includes an additive constant $e_0$:
\[
e(t) \simeq e_0 + a \cdot t^{-1/2}.
\]
The value of the constant corresponds to the average energy of the
glassy states at zero temperature \cite{deoliveira01b}. The
existence of freezing at zero temperature and the role of
activated process at low temperature was first pointed out by
Vinals and Grant. Domains are pinned by some local finite energy
barrier \cite{vinals87,deoliveira04} which cannot be overcome if
$T\simeq 0$. For low finite temperature these states present a
lifetime independent of the system size \cite{ferrero07b}, which
is determined by the predominance of the activation time with
respect to an equilibration time \cite{vinals87}.
\end{itemize}

In addition to the above temperature ranges it is seen that, in
the range just below the transition $T_q$, some  metastable states
with energy and structure similar to the disordered phase can be
observed \cite{ferrero07b,ibaneztbs}. Investigation of the nature
of these states is the subject of the next section.

\section{Metastable states}

\subsection{Theoretical premises}

Existence of metastable states just below a discontinuous
transition is predicted by phenomenological theories like Van der
Waals equation for the liquid-gas transitions \cite{gunton83} and
are a general feature of theories with mean-field or long range
interaction \cite{binder87}. As in the Van der Waals equation,
metastable branches appear as a continuation of the stable branch
beyond the transition point and have to be replaced by the
coexistence curve, e.g. via the Maxwell construction. Along these
branches the metastable state is thermodynamically stable as long
as the second free energy derivative is positive and the escape
from from the local minimum takes places via thermal activation.
The theory of Ginzburg and Landau takes into account localized
fluctuations, but these cannot be very large, excluding therefore
interesting regions from its range of validity, like the one
around the critical temperature. At the end of the metastable
branch is the spinodal point. There the second derivative of the
free energy changes of sign, making the whole system unstable. The
decay to the stable phase is global and quick.
Beyond mean field, statistical mechanics of short-interacting
systems cannot account properly for metastability, since the
partition function in ensemble theory is dominated by the global
minimum of the free energy functional in phase space.

Metastable states with infinite lifetime in short range system with
translational invariance are thermodynamically forbidden by
stability requirements \cite{griffiths64}, although they can be
obtained by setting specific constraints to the accessible phase
space or taking suitable limits in some perturbative approaches
\cite{binder87}. The constraint is related to a timescale, up to
which the metastable phase do not explore the whole available
phase space. In this context, in the droplet theory for the
condensation of the field-driven Ising transition \cite{langer67},
the fluctuations neglected in mean-field approaches give raise
locally to droplets of the stable phase. An ensemble of
non-interacting droplets of the stable phase is considered, its
growth being favored by the bulk free energy, which is lower in
presence of an external field, and hindered by the surface
tension. The model has been extensively tested in two and three
dimensions \cite{rikvold02}.

For the $T$-driven Potts model transition,
differently from what happens in the Ising case, the growth of droplets
just below $T_q$ is favoured by the tendency to lower the total
energy of the system (the interface perimeter between droplets),
to the detriment of the system entropy (which is higher when many small
droplets are present). In this case, the ensemble of droplets is
also different with respect to the Ising case in that droplets
have an intrinsic entropy, and can assume different shapes. 
Meunier and Morel \cite{meunier00} considered an adaptation of the
droplet approach to the 2D Potts model. Their theory leads to a 
temperature range in which
metastability is present in finite-size systems, range which,
however, shrinks to zero in the thermodynamic limit. This
behaviour, related in some way to the anomalous growth of
fluctuations with the system size near the critical point, has, to
our knowledge, not yet been confirmed by dynamic calculations.
Investigation of metastable states from a pure dynamical point of
view seems therefore to deserve interest.

\subsection{Some numerical results}

Numerical simulation of the Potts model with more than four states
in $2D$ constitutes a very stimulating and effective way for
investigating the nature of metastable states in systems with many
low-temperature stable phases.

A way for identifying  metastable states just below the transition
temperature is to look at the time behaviour of the  energy per
site $e(t)$. As a test system we have considered the Potts model
on square lattices with $q=12,50$. Simulations have been performed
by preparing the system in the high temperature disordered phase
and then quenching it by Metropolis dynamics below the transition
temperature $T_q$.

\begin{figure}
\epsfig{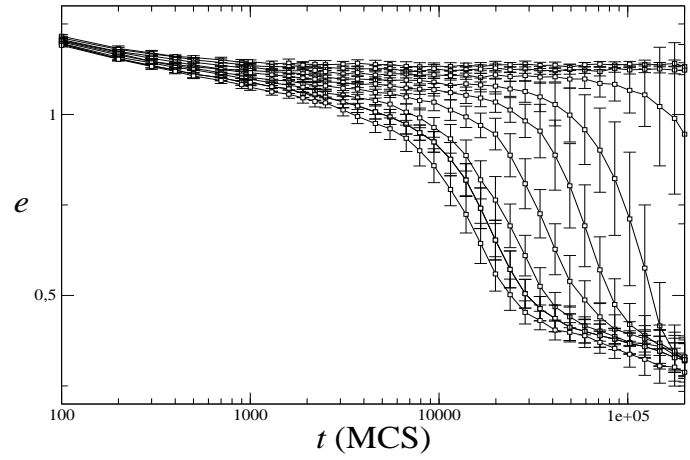}%
\caption{Time behaviour of the energy per site, after the quench
of the model with $q=12$ at different quenching temperatures below
$T_q$ (see text). The  quenching temperature increases from bottom
to top.}\label{figure1}
\end{figure}

For the $q=12$ system, for which $=T_q=1/\beta_q=0.668$, we have
taken the quench inverse temperature at $\beta=\beta_q 
(1+ j\cdot0.005$), with $j=1,\dots10$. The resulting time behaviour for
lattices of size $L=250$  is shown in figure \ref{figure1}. Each
line is obtained by averaging over $50$ different realizations of
initial conditions and thermal noise, and the errors are the
resulting standard deviations. It is seen that for temperature
close to $T_q$, $e(t)$ stays stationary (a signature of
metastability \cite{binder87}) until large times, at an excess
energy with respect to the equilibrium state. The existence of
these metastable states is in agreement with \cite{meunier00}, who
however predict their disappearance in the thermodynamic limit.

From the point of view of the local dynamics, if the system is in
a metastable state just below the transition temperature, it is
expected to be indistinguishable from one in equilibrium.  This
property is reflected in the time behaviour of the
self-correlation function. In the equilibrium disordered phase
($T>T_q$) this function displays a fast (exponential) relaxation.
The same behaviour is therefore expected for the metastable states
below $T_q$, but will no longer be true after the decay towards
the stable state has begun. The correlation function computed
waiting different times $t_w$ after the quench time should
therefore signal this decay through the onset of a slower decay
(ageing). The self-correlation function can be computed as:
\begin{equation}
\label{correlation}
 c(t,t_w)=\frac{1}{q-1} \langle(\frac{q}{N} \sum \delta_{\eta_i(t+t_w),\eta_i(t_w)}-1)\rangle,
\end{equation}
where $t_w$ is the time elapsed after the quench before starting
the computation and the sum is over all the lattice site (averages over different
realizations of initial conditions and thermal noise are taken).

\begin{figure}c
 \epsfig{file=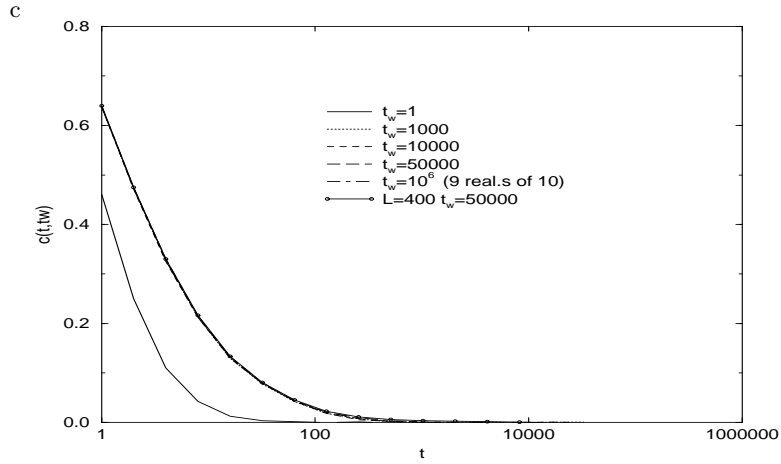, width=6cm, height=10cm, angle=-90}
\caption{Time behaviour of the self-correlation function for
different waiting times for a system with $q=50$. The lattice is
$L=100$  in size and the quench is at the temperature $T=0.470$
($T_q=0.479$)}. \label{figure2}
\end{figure}

As mentioned before, it is expected that for $t_w$ short enough
the relaxation is fast. Figure \ref{figure2} shows $c(t,t_w)$ for
different $t_w$ for a system with $q=50$, $L=100$ and $T=0.470$,
being $T_q=0.479$. Curves are obtained by averaging over $10$
realizations  (except one indicated in the legend). It is seen
that after a very short transient time $c(t,t_w)$ becomes
translationally invariant for long times, also showing no
dependence on the size in the investigated range.

\begin{figure}
\epsfig{file=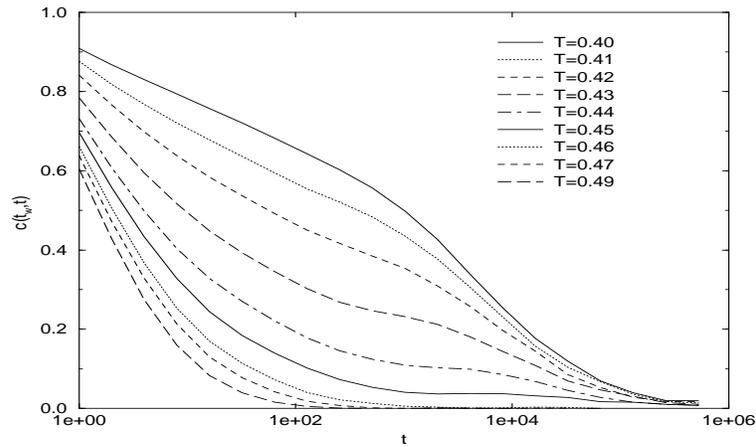, width=6cm, height=10cm, angle=-90}%
\caption{Time behaviour of the self-correlation-function for
$t_w=1000$ after quenching at different temperatures below
$T_q=0.479$. At low  temperatures ageing appears.}\label{figure3}
\end{figure}

Figure \ref{figure3} shows $c(t,t_w)$ vs  time $t$ for $q=50$,
$t_w=1000$ and different temperatures.  Ageing becomes larger when
the temperature is decreased, showing that the system is already
relaxing toward the ordering phase.

These results show the existence of well definite metastable
states on square lattices of finite size. The question of their
existence in the thermodynamic limit is thus open.  Meunier and
Morel \cite{meunier00} have recently faced the problem by
investigating the energy probability distribution at the
transition temperature, concluding that metastability is
observable only in systems of finite size. The anomalous growth of
energy fluctuations near the critical point is at the origin of
this fact, but the microscopic mechanisms behind this shrinking 
of the metastability interval for large systems remain unknown.
It would be very interesting, however, to have a description of this
phenomena in terms of microscopic clusters. One could then compare the
mechanisms of nucleation in the Potts case with those predicted for the
field-driven Ising case by Classical Nuclation Theory.
In Classical nucleation theory, when the system size is much larger than
the lenthscales involved in the the nucleation process then lifetimes and
properties of the metastable phase do not depend on the system size 
\cite{rikvold02}. What is, hence, the microscopic difference in the $T$-driven
Potts transition, with respect to this scenario, which leads to the
shrinking of the metastable temperature range for large system?

According to the nucleation theory the finite lifetime is usually
determined by the time needed to the system for nucleating a drop
of the stable phase bigger than the critical size. At the same
time, as mentioned above, the metastable state displays short
relaxation times similar to the high temperature phase. The latter
usually increase when temperature decreases, and it may happens to
become equal to the nucleation time at a certain temperature
$T_s$. This temperature corresponds to a pseudo-spinodal point
\cite{debenedetti} and sets a limit to the observability of
undercooled metastable states. The first point is thus to
establish if $T_s < T_q$ even for infinite systems or if the two
temperature merge. Another point concerns the growing dynamics
following the nucleation. In the Ising model the metastable state
nucleates a critical droplet that afterward grows very fast
because of its favourable free energy with respect to the unstable
phase. In the temperature driven Potts model however each phase is
equivalent to the others and competes with them for the growth.

\section{Conclusions}

We have discussed some aspects of the off-equilibrium dynamics of
the $q>4$ Potts model quenched below the transition temperature.
We believe that this can be relevant for understanding the
non-equilibrium and the ordering dynamics of systems below a first
order phase transition temperature, in the presence of many
competing equivalent ground states. After recalling some recent
findings we have focussed on some characterization of the
undercooled metastable states which are observable after a quench
below the transition temperature. These states display a finite
lifetime which depends on the temperature and are detectable
through the existence of a plateau in the energy (well above the
equilibrium energy) and the fast relaxation time. Being their
existence in the thermodynamical limit questioned by recent work,
more extensive numerical simulations will be necessary to support
this statement. More work in this field is highly desirable also
for understanding the detailed mechanism through which a system
starts to order after the quench in the presence of many
degenerate states, which is not yet well established.

\section*{Acknowledgments}

Part of the work was supported by the Italian Government Contract
FIRB RBAU01883Z. A.P. thanks M.J. de Oliveira for many stimulating
and useful discussions.

\bibliography{petri_2}

\begin{thebibliography}{29}
\expandafter\ifx\csname natexlab\endcsname\relax\def\natexlab#1{#1}\fi
\expandafter\ifx\csname bibnamefont\endcsname\relax
  \def\bibnamefont#1{#1}\fi
\expandafter\ifx\csname bibfnamefont\endcsname\relax
  \def\bibfnamefont#1{#1}\fi
\expandafter\ifx\csname citenamefont\endcsname\relax
  \def\citenamefont#1{#1}\fi
\expandafter\ifx\csname url\endcsname\relax
  \def\url#1{\texttt{#1}}\fi
\expandafter\ifx\csname urlprefix\endcsname\relax\def\urlprefix{URL }\fi
\providecommand{\bibinfo}[2]{#2}
\providecommand{\eprint}[2][]{\url{#2}}

\bibitem[{\citenamefont{Bray}(1994)}]{bray94}
\bibinfo{author}{\bibfnamefont{A.}~\bibnamefont{Bray}},
  \bibinfo{journal}{Advances in Physics} \textbf{\bibinfo{volume}{43}},
  \bibinfo{pages}{357} (\bibinfo{year}{1994}).

\bibitem[{\citenamefont{Hohenberg and Halperin}(1977)}]{hoenberg77}
\bibinfo{author}{\bibfnamefont{P.}~\bibnamefont{Hohenberg}} \bibnamefont{and}
  \bibinfo{author}{\bibfnamefont{B.}~\bibnamefont{Halperin}},
  \bibinfo{journal}{Rev. Mod. Phys.} \textbf{\bibinfo{volume}{49}},
  \bibinfo{pages}{435} (\bibinfo{year}{1977}).

\bibitem[{\citenamefont{Gunton et~al.}(1983)\citenamefont{Gunton, Miguel, and
  Sahni}}]{gunton83}
\bibinfo{author}{\bibfnamefont{J.}~\bibnamefont{Gunton}},
  \bibinfo{author}{\bibfnamefont{M.~S.} \bibnamefont{Miguel}},
  \bibnamefont{and} \bibinfo{author}{\bibfnamefont{P.~S.} \bibnamefont{Sahni}},
  in \emph{\bibinfo{booktitle}{Phase Transitions and Critical Phenomena}},
  edited by \bibinfo{editor}{\bibfnamefont{C.}~\bibnamefont{Domb}}
  \bibnamefont{and} \bibinfo{editor}{\bibfnamefont{M.~S.} \bibnamefont{Green}}
  (\bibinfo{publisher}{Academic, New York}, \bibinfo{year}{1983}),
  vol.~\bibinfo{volume}{8}, chap. \bibinfo{chapter}{The Dynamics of First-Order
  Phase Transition}, pp. \bibinfo{pages}{267--482}.

\bibitem[{\citenamefont{Binder}(1987)}]{binder87}
\bibinfo{author}{\bibfnamefont{K.}~\bibnamefont{Binder}},
  \bibinfo{journal}{Rep. Prog. Phys.} \textbf{\bibinfo{volume}{50}},
  \bibinfo{pages}{783} (\bibinfo{year}{1987}).

\bibitem[{\citenamefont{Langer}(1967)}]{langer67}
\bibinfo{author}{\bibfnamefont{J.~S.} \bibnamefont{Langer}},
  \bibinfo{journal}{Annals of Physics} \textbf{\bibinfo{volume}{41}},
  \bibinfo{pages}{108} (\bibinfo{year}{1967}).

\bibitem[{\citenamefont{Rikvold et~al.}(1994)\citenamefont{Rikvold, Tomita,
  Miyashita, and Sides}}]{rikvold94}
\bibinfo{author}{\bibfnamefont{P.~A.} \bibnamefont{Rikvold}},
  \bibinfo{author}{\bibfnamefont{H.}~\bibnamefont{Tomita}},
  \bibinfo{author}{\bibfnamefont{S.}~\bibnamefont{Miyashita}},
  \bibnamefont{and} \bibinfo{author}{\bibfnamefont{S.~W.} \bibnamefont{Sides}},
  \bibinfo{journal}{Physical Review E} \textbf{\bibinfo{volume}{49}},
  \bibinfo{pages}{5080 } (\bibinfo{year}{1994}).

\bibitem[{\citenamefont{Novotny et~al.}(2002)\citenamefont{Novotny, Brown, and
  Rikvold}}]{rikvold02}
\bibinfo{author}{\bibfnamefont{M.~A.} \bibnamefont{Novotny}},
  \bibinfo{author}{\bibfnamefont{G.}~\bibnamefont{Brown}}, \bibnamefont{and}
  \bibinfo{author}{\bibfnamefont{P.~A.} \bibnamefont{Rikvold}},
  \bibinfo{journal}{Journal of applied Physics} \textbf{\bibinfo{volume}{91}},
  \bibinfo{pages}{6908} (\bibinfo{year}{2002}).

\bibitem[{\citenamefont{Schmittmann and Bruce}(1985)}]{schmittmann85}
\bibinfo{author}{\bibfnamefont{B.}~\bibnamefont{Schmittmann}} \bibnamefont{and}
  \bibinfo{author}{\bibfnamefont{A.~D.} \bibnamefont{Bruce}},
  \bibinfo{journal}{J. Phys. A: Math. Gen.} \textbf{\bibinfo{volume}{18}},
  \bibinfo{pages}{1715} (\bibinfo{year}{1985}).

\bibitem[{\citenamefont{Lifshitz}(1962)}]{lifshitz62}
\bibinfo{author}{\bibfnamefont{I.}~\bibnamefont{Lifshitz}},
  \bibinfo{journal}{Zh. Eksp. Teor. Fiz.} \textbf{\bibinfo{volume}{42}},
  \bibinfo{pages}{1354} (\bibinfo{year}{1962}).

\bibitem[{\citenamefont{Meunier and Morel}(2000)}]{meunier00}
\bibinfo{author}{\bibfnamefont{J.~L.} \bibnamefont{Meunier}} \bibnamefont{and}
  \bibinfo{author}{\bibfnamefont{A.}~\bibnamefont{Morel}},
  \bibinfo{journal}{European Physical Journal B} \textbf{\bibinfo{volume}{13}},
  \bibinfo{pages}{341} (\bibinfo{year}{2000}).

\bibitem[{\citenamefont{Binder}(1981)}]{binder81}
\bibinfo{author}{\bibfnamefont{K.}~\bibnamefont{Binder}}, \bibinfo{journal}{J.
  Stat. Phys.} \textbf{\bibinfo{volume}{24}}, \bibinfo{pages}{69}
  (\bibinfo{year}{1981}).

\bibitem[{\citenamefont{Berg et~al.}(2004)\citenamefont{Berg, Heller,
  Meyer-Ortmanns, and Velytsky}}]{berg04}
\bibinfo{author}{\bibfnamefont{B.~A.} \bibnamefont{Berg}},
  \bibinfo{author}{\bibfnamefont{U.~M.} \bibnamefont{Heller}},
  \bibinfo{author}{\bibfnamefont{H.}~\bibnamefont{Meyer-Ortmanns}},
  \bibnamefont{and} \bibinfo{author}{\bibfnamefont{A.}~\bibnamefont{Velytsky}},
  \bibinfo{journal}{Phys. Rev. D} \textbf{\bibinfo{volume}{69}},
  \bibinfo{pages}{034501} (\bibinfo{year}{2004}).

\bibitem[{\citenamefont{Rutkevich}(2002)}]{rutkevich02}
\bibinfo{author}{\bibfnamefont{S.~B.} \bibnamefont{Rutkevich}},
  \bibinfo{journal}{International Journal of Modern Physics C}
  \textbf{\bibinfo{volume}{13}}, \bibinfo{pages}{495} (\bibinfo{year}{2002}).

\bibitem[{\citenamefont{Wu}(1982)}]{wu82}
\bibinfo{author}{\bibfnamefont{I.}~\bibnamefont{Wu}}, \bibinfo{journal}{Rev.
  Mod. Phys} \textbf{\bibinfo{volume}{54}}, \bibinfo{pages}{235}
  (\bibinfo{year}{1982}).

\bibitem[{\citenamefont{Grest et~al.}(1988)\citenamefont{Grest, Anderson, and
  Srolovitz}}]{grest88}
\bibinfo{author}{\bibfnamefont{G.~S.} \bibnamefont{Grest}},
  \bibinfo{author}{\bibfnamefont{M.~P.} \bibnamefont{Anderson}},
  \bibnamefont{and} \bibinfo{author}{\bibfnamefont{D.~J.}
  \bibnamefont{Srolovitz}}, \bibinfo{journal}{Phys. Rev. B}
  \textbf{\bibinfo{volume}{38}}, \bibinfo{pages}{4752} (\bibinfo{year}{1988}).

\bibitem[{\citenamefont{Thomas et~al.}(2006)\citenamefont{Thomas, {de Almeida},
  and Graner}}]{thomas06}
\bibinfo{author}{\bibfnamefont{G.~L.} \bibnamefont{Thomas}},
  \bibinfo{author}{\bibfnamefont{R.~M.~C.} \bibnamefont{{de Almeida}}},
  \bibnamefont{and} \bibinfo{author}{\bibfnamefont{F.}~\bibnamefont{Graner}},
  \bibinfo{journal}{Physical Review E (Statistical, Nonlinear, and Soft Matter
  Physics)} \textbf{\bibinfo{volume}{74}}, \bibinfo{eid}{021407}
  (pages~\bibinfo{numpages}{18}) (\bibinfo{year}{2006}),
  \urlprefix\url{http://link.aps.org/abstract/PRE/v74/e021407}.

\bibitem[{\citenamefont{Anderson et~al.}(1989)\citenamefont{Anderson, Grest,
  and Srolovitz}}]{grest89}
\bibinfo{author}{\bibfnamefont{M.~P.} \bibnamefont{Anderson}},
  \bibinfo{author}{\bibfnamefont{G.~S.} \bibnamefont{Grest}}, \bibnamefont{and}
  \bibinfo{author}{\bibfnamefont{D.~J.} \bibnamefont{Srolovitz}},
  \bibinfo{journal}{Philosophical Magazine Part B}
  \textbf{\bibinfo{volume}{59}}, \bibinfo{pages}{293} (\bibinfo{year}{1989}).

\bibitem[{\citenamefont{Graner and Glazier}(1992)}]{graner92}
\bibinfo{author}{\bibfnamefont{F.}~\bibnamefont{Graner}} \bibnamefont{and}
  \bibinfo{author}{\bibfnamefont{J.~A.} \bibnamefont{Glazier}},
  \bibinfo{journal}{Phys. Rev. Lett.} \textbf{\bibinfo{volume}{69}},
  \bibinfo{pages}{2013} (\bibinfo{year}{1992}).

\bibitem[{\citenamefont{Derrida et~al.}(1996)\citenamefont{Derrida, {de
  Oliveira}, and Stauffer}}]{derrida96}
\bibinfo{author}{\bibfnamefont{B.}~\bibnamefont{Derrida}},
  \bibinfo{author}{\bibfnamefont{P.~M.~C.} \bibnamefont{{de Oliveira}}},
  \bibnamefont{and} \bibinfo{author}{\bibfnamefont{D.}~\bibnamefont{Stauffer}},
  \bibinfo{journal}{Physica A: Statistical and Theoretical Physics}
  \textbf{\bibinfo{volume}{224}}, \bibinfo{pages}{604} (\bibinfo{year}{1996}).

\bibitem[{\citenamefont{{de Oliveira} and Petri}(2002)}]{deoliveira01}
\bibinfo{author}{\bibfnamefont{M.~J.} \bibnamefont{{de Oliveira}}}
  \bibnamefont{and} \bibinfo{author}{\bibfnamefont{A.}~\bibnamefont{Petri}},
  \bibinfo{journal}{Phil. Mag. B} pp. \bibinfo{pages}{617--623}
  (\bibinfo{year}{2002}).

\bibitem[{\citenamefont{{Ib\'a\~nez de Berganza}
  et~al.}(2007{\natexlab{a}})\citenamefont{{Ib\'a\~nez de Berganza}, Ferrero,
  Cannas, Loreto, and Petri}}]{ferrero07}
\bibinfo{author}{\bibfnamefont{M.}~\bibnamefont{{Ib\'a\~nez de Berganza}}},
  \bibinfo{author}{\bibfnamefont{E.~E.} \bibnamefont{Ferrero}},
  \bibinfo{author}{\bibfnamefont{S.~A.} \bibnamefont{Cannas}},
  \bibinfo{author}{\bibfnamefont{V.}~\bibnamefont{Loreto}}, \bibnamefont{and}
  \bibinfo{author}{\bibfnamefont{A.}~\bibnamefont{Petri}},
  \bibinfo{journal}{Euro Physical Journal, Special Topics}
  \textbf{\bibinfo{volume}{143}} (\bibinfo{year}{2007}{\natexlab{a}}).

\bibitem[{\citenamefont{Ferrero and Cannas}(2007)}]{ferrero07b}
\bibinfo{author}{\bibfnamefont{E.}~\bibnamefont{Ferrero}} \bibnamefont{and}
  \bibinfo{author}{\bibfnamefont{S.}~\bibnamefont{Cannas}},
  \bibinfo{journal}{Phys. Rev. E} \textbf{\bibinfo{volume}{76}},
  \bibinfo{pages}{031108} (\bibinfo{year}{2007}).

\bibitem[{\citenamefont{{Ib\'a\~nez de Berganza}
  et~al.}(2007{\natexlab{b}})\citenamefont{{Ib\'a\~nez de Berganza}, V.Loreto,
  and A.Petri}}]{ibanez07}
\bibinfo{author}{\bibfnamefont{M.}~\bibnamefont{{Ib\'a\~nez de Berganza}}},
  \bibinfo{author}{\bibnamefont{V.Loreto}}, \bibnamefont{and}
  \bibinfo{author}{\bibnamefont{A.Petri}}, \bibinfo{journal}{Philosophical
  Magazine B} \textbf{\bibinfo{volume}{87}}, \bibinfo{pages}{779}
  (\bibinfo{year}{2007}{\natexlab{b}}).

\bibitem[{\citenamefont{{de Oliveira}
  et~al.}(2004{\natexlab{a}})\citenamefont{{de Oliveira}, Petri, and
  Tom\'e}}]{deoliveira01b}
\bibinfo{author}{\bibfnamefont{M.~J.} \bibnamefont{{de Oliveira}}},
  \bibinfo{author}{\bibfnamefont{A.}~\bibnamefont{Petri}}, \bibnamefont{and}
  \bibinfo{author}{\bibfnamefont{T.}~\bibnamefont{Tom\'e}},
  \bibinfo{journal}{Europhys. Lett.}  (\bibinfo{year}{2004}{\natexlab{a}}).

\bibitem[{\citenamefont{Vinals and Grant}(1987)}]{vinals87}
\bibinfo{author}{\bibfnamefont{J.}~\bibnamefont{Vinals}} \bibnamefont{and}
  \bibinfo{author}{\bibfnamefont{M.}~\bibnamefont{Grant}},
  \bibinfo{journal}{Phys. Rev. B} \textbf{\bibinfo{volume}{36}},
  \bibinfo{pages}{7036} (\bibinfo{year}{1987}).

\bibitem[{\citenamefont{{de Oliveira}
  et~al.}(2004{\natexlab{b}})\citenamefont{{de Oliveira}, Petri, and
  Tom\'e}}]{deoliveira04}
\bibinfo{author}{\bibfnamefont{M.~J.} \bibnamefont{{de Oliveira}}},
  \bibinfo{author}{\bibfnamefont{A.}~\bibnamefont{Petri}}, \bibnamefont{and}
  \bibinfo{author}{\bibfnamefont{T.}~\bibnamefont{Tom\'e}},
  \bibinfo{journal}{Physica A} \textbf{\bibinfo{volume}{342}},
  \bibinfo{pages}{97} (\bibinfo{year}{2004}{\natexlab{b}}).

\bibitem[{\citenamefont{{Ib\'a\~nez de Berganza}
  et~al.}(2008)\citenamefont{{Ib\'a\~nez de Berganza}, Loreto, and
  Petri}}]{ibaneztbs}
\bibinfo{author}{\bibfnamefont{M.}~\bibnamefont{{Ib\'a\~nez de Berganza}}},
  \bibinfo{author}{\bibfnamefont{V.}~\bibnamefont{Loreto}}, \bibnamefont{and}
  \bibinfo{author}{\bibfnamefont{A.}~\bibnamefont{Petri}},
  \bibinfo{journal}{arXiv:0706.3534v1, Submitted}  (\bibinfo{year}{2008}).

\bibitem[{\citenamefont{Griffiths}(1964)}]{griffiths64}
\bibinfo{author}{\bibfnamefont{R.~B.} \bibnamefont{Griffiths}},
  \bibinfo{journal}{J. Math. Phys.} \textbf{\bibinfo{volume}{5}},
  \bibinfo{pages}{1215} (\bibinfo{year}{1964}).

\bibitem[{\citenamefont{Debenedetti}(1996)}]{debenedetti}
\bibinfo{author}{\bibfnamefont{P.~G.} \bibnamefont{Debenedetti}},
  \emph{\bibinfo{title}{Metastable Liquids: Concepts and Principles}}
  (\bibinfo{publisher}{Princeton University Press, Princeton},
  \bibinfo{year}{1996}).

\end{thebibliography}

\newpage

\vspace{5cm}

\end{document}